\documentclass[aps,twocolumn,prl,superscriptaddress,amsmath]{revtex4}
\usepackage{epsfig,graphicx,times}
\usepackage{amstext}
\usepackage{amsmath}
\usepackage{amssymb}
\usepackage{graphicx}
\usepackage{latexsym}
\usepackage{bm}
\usepackage[UKenglish]{babel}
\usepackage[normalem]{ulem}

\begin{document}

\title{Tuneable on-demand single-photon source}

\author{Z.H. Peng}\email{zhihui_peng@riken.jp}
\affiliation{Physics Department, Royal Holloway, University of London, Egham, Surrey TW20 0EX, United Kingdom}\affiliation{Center for Emergent Matter Science, RIKEN, Wako, Saitama 351-0198, Japan}
\author{J.S. Tsai}
\affiliation{Department of Physics, Tokyo University of Science, Kagurazaka, Tokyo 162-8601, Japan}
\affiliation{Center for Emergent Matter Science, RIKEN, Wako, Saitama 351-0198, Japan}
\author{O.V. Astafiev}\email{Oleg.Astafiev@rhul.ac.uk}
\affiliation{Physics Department, Royal Holloway, University of London, Egham, Surrey TW20 0EX, United Kingdom}\affiliation{National Physical Laboratory, Teddington, TW11 0LW, United Kindom}\affiliation{Moscow Institute of Physics and Technology, Dolgoprudny, 141700, Russia}

\begin{abstract}
An on-demand single photon source is a key element in a series of prospective quantum technologies and applications. We demonstrate the operation of a tuneable on-demand microwave photon source based on a fully controllable superconducting artificial atom strongly coupled to an open-end transmission line (a 1D half-space). The atom emits a photon upon excitation by a short microwave $\pi$-pulse applied through a control line weakly coupled to the atom. The emission and control lines are well decoupled from each other, preventing the direct leakage of radiation from the $\pi$-pulses used for excitation. The estimated efficiency of the source is higher than 75\% and remains to be about 50\% or higher over a wide frequency range from 6.7 to 9.1 GHz continuously tuned by an external magnetic field.
\end{abstract}
\maketitle
Control and manipulation with light at the single-photon level \cite{Kimble2008,Duan2010,Sangouard2011,Northup2014} is interesting from fundamental and practical viewpoints. In particular, on-demand single-photon sources are of high interest due to their promising applications in quantum communication, quantum informatics, sensing, and other fields. In spite of several realisations in optics \cite{Kim1999,Lounis2000,Kurtsiefer2000,Keller2004,He2013}, practical implementation of the photon sources imposes a number of requirements, such as high photon generation and collection efficiencies, frequency tunability. Recently developed superconducting quantum systems provide a novel basis for the realisation of microwave (MW) photon sources with photons confined in resonator modes  \cite{Houck2006,Hofheinz2008,Bozyigit2011,Lang2011,Lang2013,Pechal2014}. However, all the circuits consist of two elements (the resonator and the quantum emitter system) and the generated photon frequency fixed to the resonator frequency. We propose and realise a different approach: a single-photon source based on a tuneable artificial atom coupled asymmetrically to two open-end transmission lines (1D half-spaces). The atom is excited from a weakly coupled control line ($c$) and emits a photon to a strongly coupled emission line ($e$). The photon freely propagates in the emission line and can be further processed using, for example, nonlinear circuit elements. Among the advantages of our circuit is its simplicity: it consists of the single element.

\begin{figure}
\center
\includegraphics[scale=0.5]{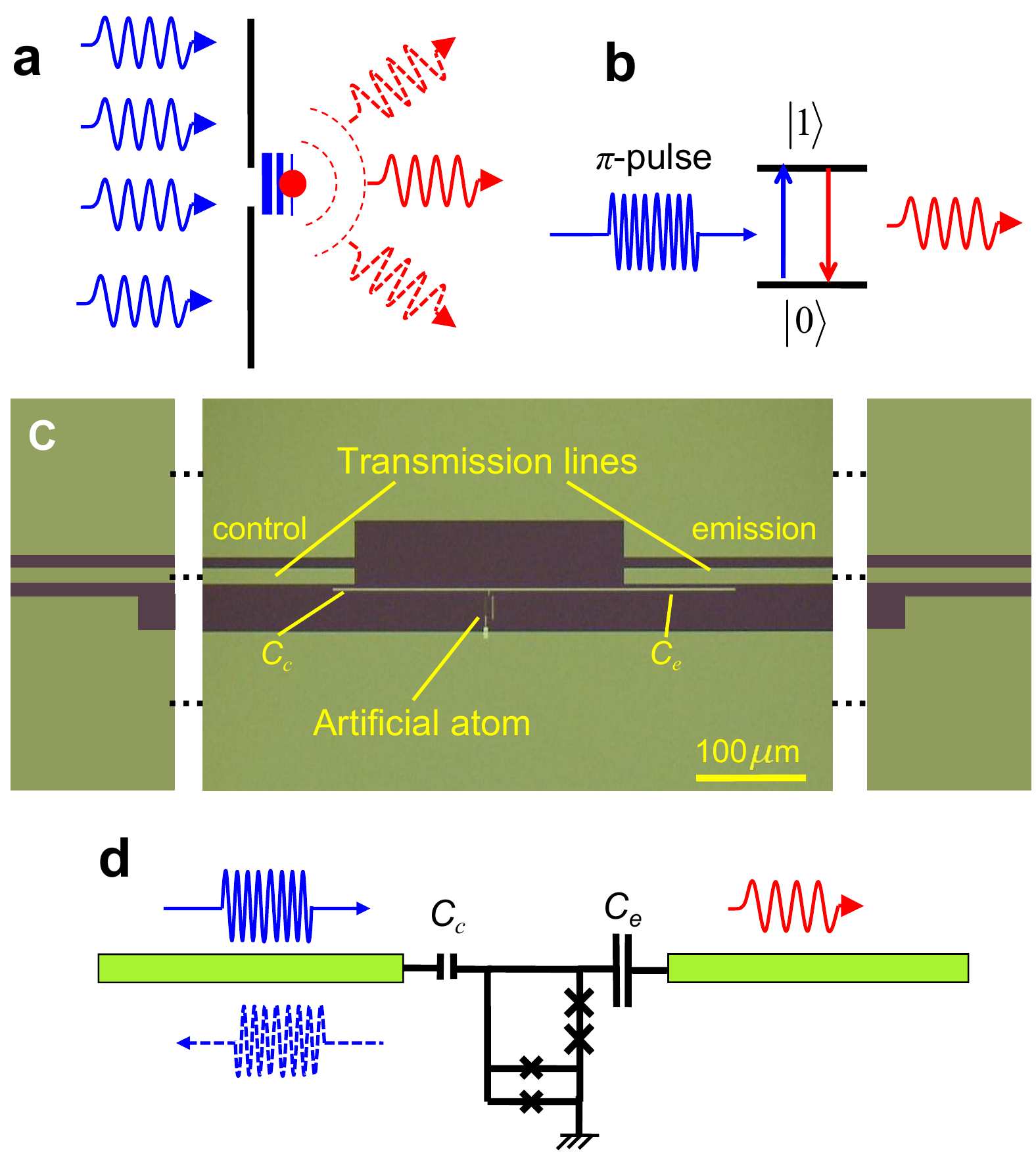}
\caption{{\bf} Single-photon source. {(\bf a)} Optical analogue of the source. {(\bf b)} Mechanism of single-photon generation. The reemitted single photon (red) is decoupled from the incident $\pi$-pulse (blue), which prepares state $|1\rangle$ in the atom. {(\bf c)} The optical micrograph of the device. The artificial atom is in the middle and the thin long metallic line from the atomic loop forms capacitances between the atom and the control/emission transmission lines. The left and right panels show transition from the central parts of lines to the usual coplanar ones. {(\bf d)} Equivalent electrical circuit of the photon source.}
\label{devicelayout}
\end{figure}

\noindent \textbf{Operation principles and device description.}
An optical analogue of the proposed single-photon source consists of a two-level atom situated near a tiny hole (much smaller than the wavelength) in a non-transparent screen (Fig.~\ref{devicelayout}a). The atom is slightly shifted towards the right-hand-side space, defining an asymmetric coupling to the half-spaces. By applying a powerful light from the left side, the atom can be excited by evanescent waves, which cannot propagate in the right-hand-side space due to their rapid decay. On the other hand, the excited atom emits photons into the right-hand-side space (Fig.~\ref{devicelayout}b). In practice, the presented layout is difficult to build using natural atoms and, even if one succeeds, another problem must be solved: the low collection efficiency of emitted photons in the 3D space.
These problems can be easily avoided by using on-chip superconducting quantum circuits coupled to 1D transmission lines \cite{Astafiev2010,Abdumalikov2011,Hoi2012}. Fig.~\ref{devicelayout}c shows a circuit with an artificial atom coupled asymmetrically to a pair of open-end coplanar transmission lines (1D-half spaces), each with $Z$ = 50~$\Omega$ impedance. The coupling capacitances  $C_c$ and $C_e$ are between the artificial atom and the control and emission lines, respectively  (shown on the equivalent circuit in Fig.~\ref{devicelayout}d). The capacitances can be approximated as point-like objects because their sizes are much smaller than the wavelength of the radiation ($\sim$ 1~cm). Note also that the transmission lines in the centre of our device (about 240~$\mu$m for each line) slightly differ from 50~$\Omega$ due to the shifted down ground plane, however we can ignore it because they are also much shorter than the wavelength.
A MW pulse is applied from the control line, exciting the atom, and then the atom emits a photon mainly to the emission line due to asymmetric coupling: $C_e / C_c \approx$~5. The following are intrinsic features of the device: (i) The two lines are well isolated from each other so that the excitation pulse does not leak from the control line to the emission line; (ii) due to the strong asymmetry, the excited atom emits a photon with up to $1 - (C_c/C_e)^2$ probability; (iii) the photon is confined in the 1D transmission line and can be easily delivered to other circuit elements through the line.

The artificial atom schematically shown in Fig.~\ref{devicelayout}d is a controllable two-level system based on a tuneable gap flux qubit \cite{Mooij1999,Wal2000,Paauw2009,Zhu2010}, that is coupled to two Nb coplanar lines. The atom is fabricated by Al/AlOx/Al shadow evaporation techniques. It contains two identical junctions in series implemented in the loop together with a dc-SQUID (called also an $\alpha$-loop), shown in the bottom part of the device in Fig. 1d. Here $\alpha\approx0.7$ specifies the nominal ratio between the two critical currents in the dc-SQUID and the other two Josephson junctions in the loop. The magnetic fluxes are quantised in the loop: an integer number, $N$, of the magnetic flux quanta, $\Phi_0$, can be trapped. At the magnetic fields where the induced magnetic flux in the loop is equal to $\Phi = \Phi_0 (N+1/2)$, two adjacent flux states $|0\rangle$ and $|1\rangle$ with $N$ and $N+1$ flux quanta, which is corresponding to oppositely circulating persistent currents, are degenerated. The degeneracy is lifted due to the finite flux tunnelling energy $\Delta_N$, determined by the effective dc-SQUID Josephson energy and varies between different degeneracy points (depends on $N$). The energy splitting of the atom $\hbar\omega_{10}=\sqrt{(2I_{p}\delta\Phi)^{2}+\Delta_N^{2}}$ is controlled by fine adjustment of the magnetic field $\delta\Phi$ in the vicinity of the degeneracy points, where $\delta\Phi=\Phi-(N+1/2)\Phi_0$ and $I_{p}$ is the persistent current in the main loop. (We neglect the weak dependence of $\Delta_N$ on $\delta\Phi$.)

\begin{figure*}
\center
\includegraphics[scale=0.8]{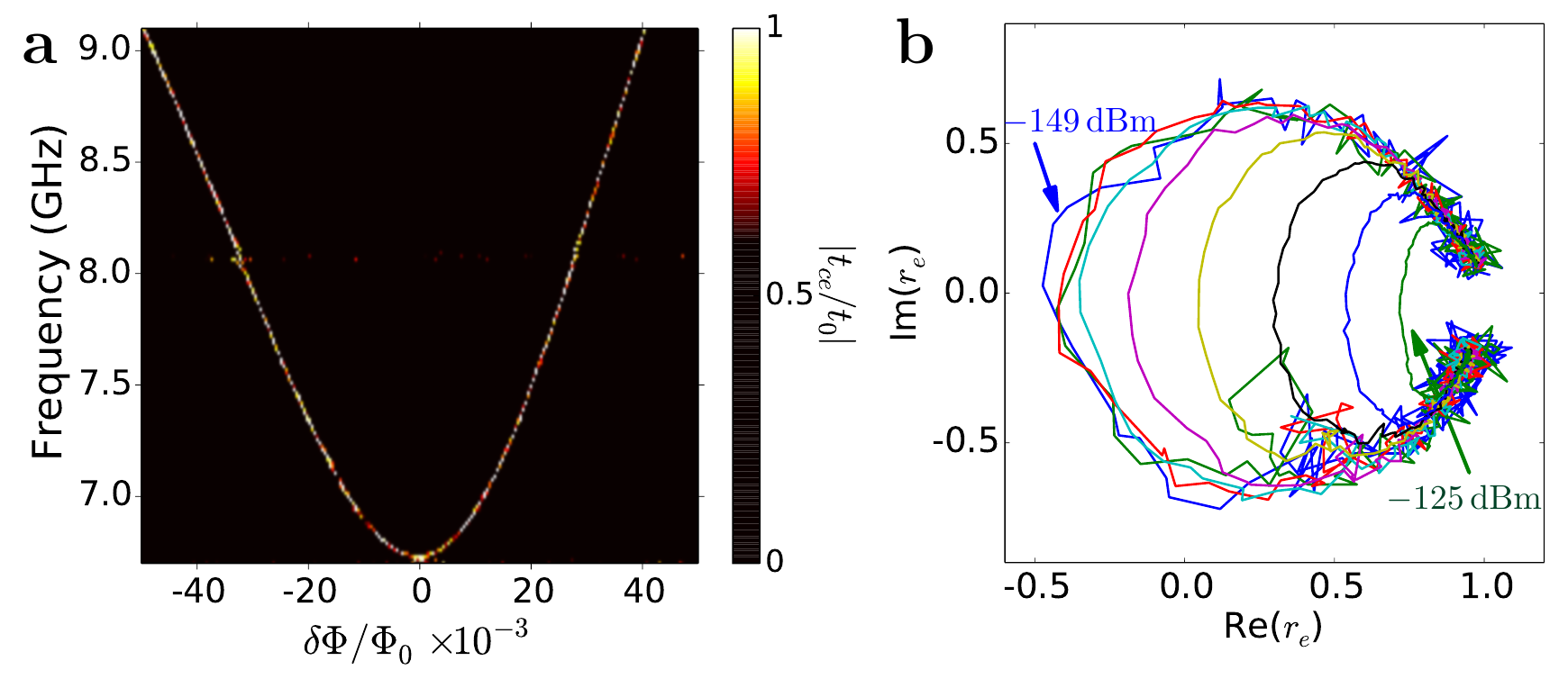}
\caption{{\bf} Transmission and reflection through the artificial atom. {(\bf a)} Transmission spectrum of the two-level artificial atom. The normalised transmission amplitude $|t_{ce}/t_{0}|$ from the control line to the emission line versus biased flux $\delta\Phi$. The transmission is suppressed everywhere except for at the resonance frequency of the atom. {(\bf b)} Smith chart of the MW reflection coefficient $r_e$ in the emission line around $\delta\Phi=0$. The reflection coefficient $r_e$ in the emission line is plotted with real-imaginary coordinates for probing powers from -149 to -125~dBm with a step of 3~dB.}
\label{transmissionspectrum}
\end{figure*}

The capacitances of the circuit are estimated to be $C_c\approx 1$~fF and $C_e\approx 5$~fF. The effective impedance between the two lines due to the capacitive coupling is $Z_{C} = 1/i\omega (C_{c}+C_e)$, which is about 3 k$\Omega$ for $\omega/2\pi$ = 10 GHz and the transmitted part of the power as low as $|2 Z/Z_{C}|^2 \approx 10^{-3}$ enables nearly perfect line decoupling.
\\
\\
\noindent \textbf{Device characterisation.}
Our experiment is carried out in a dilution refrigerator at a base temperature of around $30\,$mK. We first characterise our device by measuring the transmission coefficient $t_{ce}$ from the control line to the emission line using a vector network analyser (VNA) and the reflection coefficient $r_e$ from the emission line. Fig.~\ref{transmissionspectrum}a shows a 2D plot of the normalised transmission amplitude $|t_{ce}/t_{0}|$ in the frequency range 6.5 -- 9.1 GHz with the magnetic flux bias $\delta\Phi$ from -50 to 50 m$\Phi_0$ around the energy minimum, where $t_0$ is the maximal transmission amplitude. The transmission is suppressed everywhere except in the narrow line that corresponds to the expected atomic resonance at $\omega_{10}$ and is a result of the photon emission from the continuously driven atom. The spectroscopic curve is slightly asymmetric with respect to $\delta\Phi = 0$ due to the weak dependence of $\Delta_N$ on $\delta\Phi$. From the spectroscopy line we deduce the parameters of the two-level system: the tunnelling energy is $\Delta = \min(\hbar\omega_{10}) = h\times6.728\,$GHz at $\delta\Phi=0$ and the persistent current in the loop is $I_p \approx $ 24 nA.

To evaluate the coupling of our atom to the emission line,  we also measure the reflection at $\delta\Phi=0$ with different probing powers from -149 to -125 dBm. Next, Fig.~\ref{transmissionspectrum}b shows the reflection coefficient $r_e$ mapped on a Smith chart measured in the case of atom excitation from the emission line (opposite to the case of source operation). The curve changes its form from circular to oval, which reflects the transition from the linear weak-driving regime up to the nonlinear strong-driving regime of the two-level system~\cite{Astafiev2010}.

We next derive the dynamics of the point like atom (the loop size  $\sim10\,\rm{\mu m}$ is much smaller than the wavelength $\lambda\sim1\,$cm) located at $x = 0$ and coupled to the 1D open space via an electrical dipole. We also take into account the fact that $\omega Z C_{c,e} \ll 1$. The atom is driven by the oscillating voltage at the frequency $\omega$ of the incident wave $V_0(x,t) = V_{0}e^{-i\omega t+ikx}$ in the control line and the resulting driving amplitude of $2 V_0 \cos\omega t = {\rm{Re}}[V_0 e^{-i\omega t+ikx} + V_0 e^{-i\omega t-ikx}] |_{x=0}$ is the sum of the incident and reflected waves. The Hamiltonian of the atom in the rotating-wave approximation is $H=-(\hbar\delta\omega\sigma_{z}+\hbar\Omega\sigma_{x})/2$, where $\delta\omega=\omega-\omega_{10}$ and $\hbar\Omega = - 2V_0 C_c \nu_a$ with the electric dipole moment of the atom $\nu_a$ (between $C_c$ and $C_e$). The atomic voltage creation/annihilation operator is $\hat{\nu}^\pm = \nu_a\sigma^\pm$, where $\sigma^{\pm} = (\sigma_x \mp i\sigma_y)/2$. The driven atom generates voltage amplitudes of $V_{c,e}(t)/2 = i\omega Z C_{c,e} \nu_a \langle \sigma^- \rangle e^{-i\omega t}$ in the control ($x < 0$) and emission ($x > 0$) lines. Substituting the relaxation rates $\Gamma_1^{c,e} = S_V(\omega) (C_{c,e} \nu_a)^2/\hbar^2$ due to voltage quantum noise ($S_V(\omega) = 2\hbar\omega Z$) from the line impedance $Z$ in each line, we obtain
\begin{subequations}\label{coherentwaves}
\begin{eqnarray}
&x < 0: &V_{c}(x,t)=i\frac{\hbar\Gamma_1^{c}}{C_{c} \nu_{a}}\langle\sigma^{-}\rangle e^{-i\omega t - ikx}, \\
&x > 0:  &V_{e}(x,t)=i\frac{\hbar\Gamma_1^{e}}{C_{e} \nu_{a}}\langle\sigma^{-}\rangle e^{-i\omega t + ikx}.
\end{eqnarray}
\end{subequations}
In the ideal case of suppressed pure dephasing ($\gamma = 0$) and in the absence of nonradiative decay $\Gamma_1^{nr} = 0$, the power ratio between the control and emission lines generated by the atom under resonance is $|V_c(0,t)/V_e(0,t)|^2 = C_c^2/C_e^2 \approx 0.04$, which means that up to 96\% of the power generated by the atom can be emitted into the emission line. This allows us to measure the spectroscopy curve shown in Fig.~\ref{transmissionspectrum}a.
To find $\langle \sigma^- \rangle$ under continuous driving, we solve the master equation by considering the total relaxation rate $\Gamma_1 = \Gamma_1^c + \Gamma_1^e + \Gamma_1^{nr}$, where $\Gamma_1^{nr}$ is the nonradiative relaxation rate (for a photon absorbed by the environment). Here, the dephasing rate is $\Gamma_2 = \Gamma_1/2+\gamma$, where $\gamma$ is the pure dephasing rate. The solution is $\langle \sigma^- \rangle = -i\frac{\Omega}{2\Gamma_2}\frac{1+i\delta\omega/\Gamma_2}{1+(\delta\omega/\Gamma_2)^2+\Omega^2/\Gamma_1\Gamma_2}$. The reflection in the control line and the transmission coefficient from the control line to the emission line are $r_{c}=1+V_c(0,t)/V_0(0,t)$ and $t_{ce}=V_e(0,t)/V_0(0,t)$, respectively. At the weak-driving limit ($\Omega \ll (\Gamma_1, \Gamma_2)$),
\begin{subequations}\label{rt}
\begin{eqnarray}
&r_c& = 1-2\frac{\Gamma_1^c}{2\Gamma_2}\frac{1}{1-i\delta\omega/\Gamma_2}\label{r}, \\
&t_{ce}& = -2\frac{\Gamma_1^e}{2\Gamma_2}\frac{C_c}{C_e}\frac{1}{1-i\delta\omega/\Gamma_2}\label{t}
\end{eqnarray}
are circular plots on the Smith chart. Similarly to Eq.~(\ref{r}), we can write down the following expression for the reflection in the emission line
\begin{eqnarray}
&r_e& = 1-2\frac{\Gamma_1^e}{2\Gamma_2}\frac{1}{1-i\delta\omega/\Gamma_2},\label{re}
\end{eqnarray}
\end{subequations}
with the substitution $\Gamma_1^c$ by $\Gamma_1^e$.
We will further use this expression to characterise the coupling strength and efficiency of our device.

\begin{figure*}
\center
\includegraphics[scale=0.8]{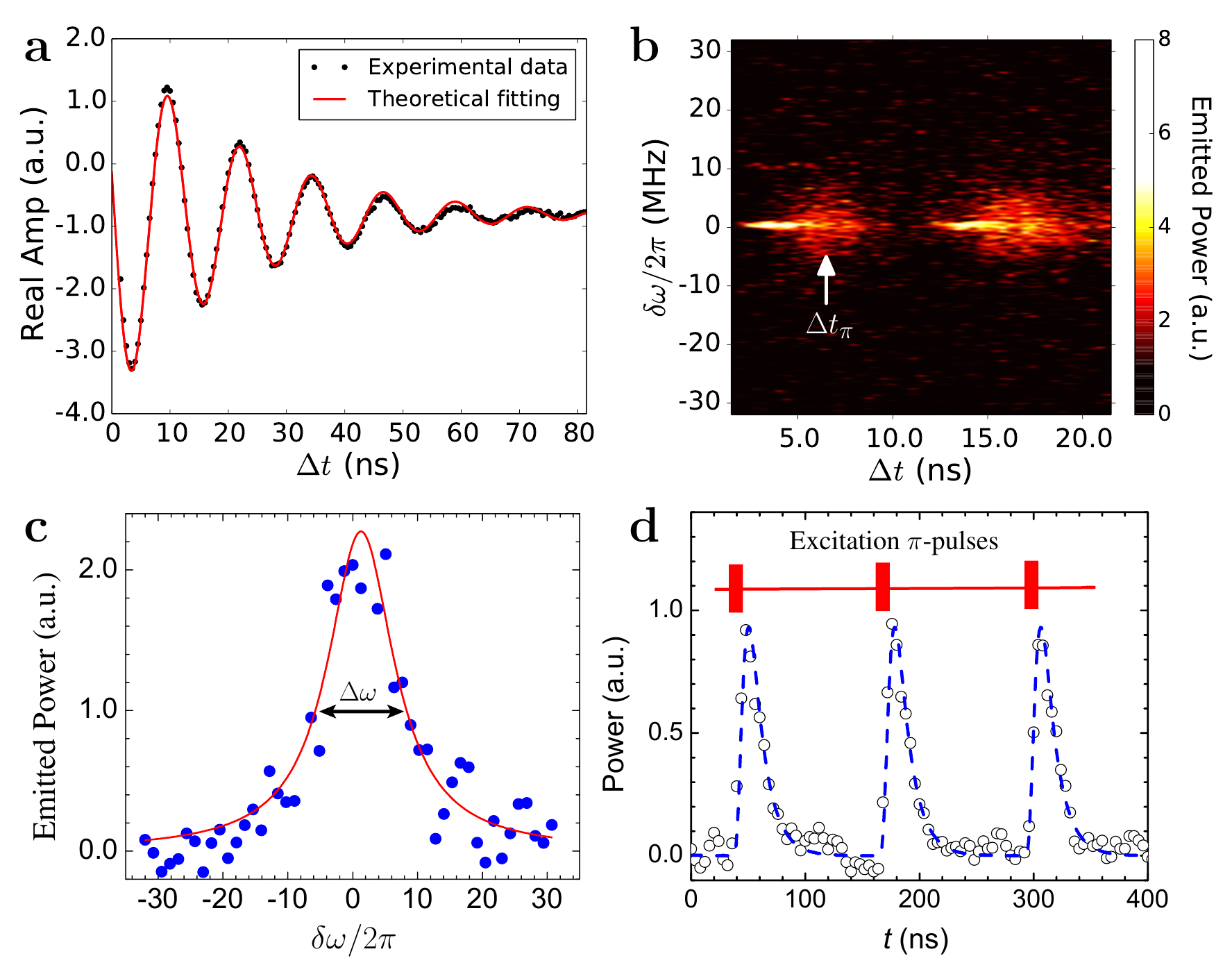}
\caption{ Single-photon source. {(\bf a)} Rabi oscillations in the two-level atom coupled to the two half spaces measured by VNA. The atom is excited by MW pulses of length $\Delta t$ from the control line with repetition time $T = 100$~ns, and the coherent emission is detected from the emission line. {(\bf b)} Emission spectrum measured by SA from the atom with the same excitation as in {\bf a}. The incoherent emission is maximal at the $\pi$-pulse ($\Delta t_\pi = 6.5$~ns). {(\bf c)} Emission peak at $\Delta t_\pi$ fitted by Lorentzian curve with FWHM $\Delta\omega$. {(\bf d)} Emission power versus time when excitation $\pi$-pulses are applied. The dashed blue line is the expected shape of the pulses with $\Gamma_1 = \Delta\omega$.}
\label{emission}
\end{figure*}

On the other hand, the excited atom emits an instantaneous power proportional to the atomic population $P_1(t)$ \cite{Abdumalikov2011} that can be straightforwardly expressed as
\begin{equation}\label{power emission}
W_{c,e}(t)=\hbar\omega\Gamma_1^{c,e}P_1(t),
\end{equation}
where $P_1 = (1-\langle\sigma_z\rangle)/2$. If the atom is prepared in the excited state $|1\rangle$ at $t = 0$, the probability decays according with $P_1(t) = \exp(-\Gamma_1 t)$.
Also, the efficiency of the photon emission to the right line is $\Gamma_1^e/\Gamma_1$, which again can be as high as $C_e^2/(C_e^2 + C_c^2) \approx 0.96$ in the ideal case.
The plot in Fig.~\ref{transmissionspectrum}b gives us a measure of the coupling strength of the atom to the emission line. Using Eq.~(\ref{re}), we estimate the efficiency of photon generation to be $\Gamma_1^e/\Gamma_1 \geq \Gamma_1^e/2\Gamma_2$ = 0.75.
\\
\noindent \textbf{Device operation.}
Our photon source based on the conversion of atomic excitation into a MW photon requires efficient control of the quantum states. Fig.~\ref{emission}a shows measured quantum oscillations. We monitor the coherent emission from the atom into the emission line by VNA when a train of identical excitation MW pulses, each of length $\Delta t$ with period $T$ = 100~ns, is applied from the control line. The amplitude of the emission oscillates with $\Delta t$. The maxima and minima of the oscillations correspond to $|\langle\sigma^\pm\rangle| \approx \pm 1$ when the atom is in the maximally superposed states with 50\% population. We also measure incoherent emission spectra using a spectrum analyser (SA) as a function of $\Delta t$. In Fig.~\ref{emission}b, one can see the central narrow peak (yellow colour) corresponding to the coherent emission ($\propto\langle\sigma^\pm\rangle$) and the wider (incoherent) emission peak ($\propto\langle\sigma_z\rangle$). The two peaks oscillate in amplitude with a $\pi/2$ shift with respect to each other, as expected from the dynamics of $\langle\sigma^\pm\rangle$ and $\langle\sigma_z\rangle$ \cite{Abdumalikov2011}. For the single photon source operation, we tune the pulse length to obtain the maximum incoherent emission (defined as a $\pi$-pulse and its length is $\Delta t_\pi$ = 6.5~ns), emitting a single photon from atom excited state in a pulse period $T$. The average emission peak excited by the $\pi$-pulses with repetition time $T = 100$~ns, which is much greater than the atom relaxation time, is shown in Fig.~\ref{emission}c. Using a Lorentzian fit, we obtain FWHM $\Delta\omega/2\pi \approx$ 12.5~MHz, which is equal to the relaxation rate $\Gamma_1$ when $\gamma = 0$.

Next, we demonstrate the operation of the photon source in time domain. We apply a train of excitation $\pi$-pulses and digitise the amplified outcoming signal from the source using a fast analog-digital converter (ADC) with a 4~ns step. To cancel the background noise from the amplifier we also take the idle traces (without applying pulses). The on-off pulse signals are squared and subtracted. Fig.~\ref{emission}d demonstrates the result of $5.12\times10^8$ times averaging. The measured peaks well coincide with the simulated ones with decay $\Gamma_1$ and 30~MHz bandwidth set by the digital filter during the data acquisition.

\begin{figure}
\center
\includegraphics[scale=0.3]{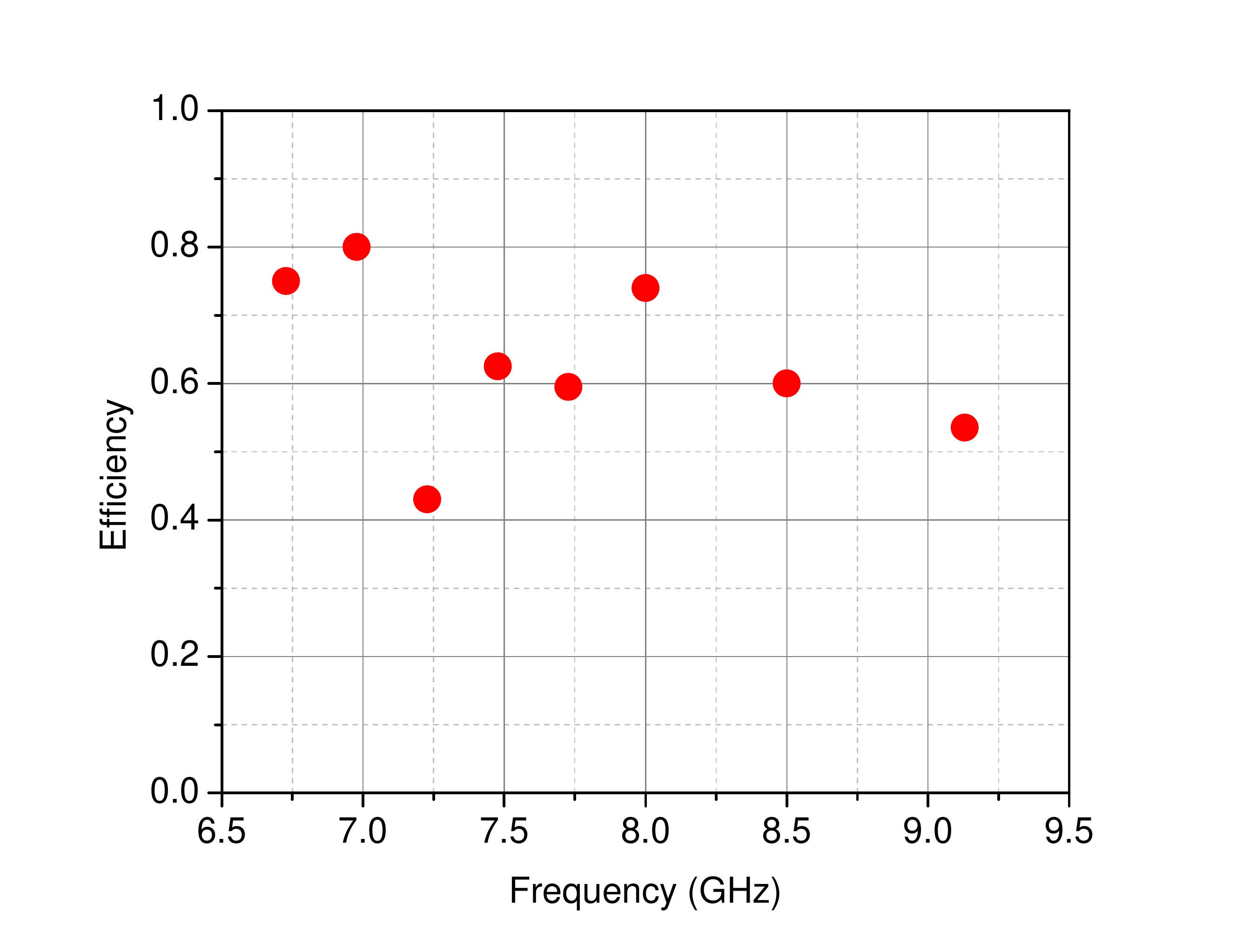}
\caption{{\bf} Extracted efficiency of the photon source versus different transition frequencies of the two-level atom. }
\label{efficiency}
\end{figure}

Finally, we evaluate the efficiency of our source over a wide frequency range by tuning the emission frequency $\omega_{10}$ controlled by $\delta\Phi$. First, we would like to point out that in our flux-qubit-based atom, pure dephasing ($\gamma \propto I_p^2$) is expected to be strongly suppressed due to the low persistent current $I_p$ being one order lower than in the conventional design \cite{Wal2000,Paauw2009}. Therefore, the pure dephasing should not affect the efficiency too much, even when we detune the energy from the minimal ones. We characterise the coupling strength using Eq.~(\ref{rt}c) by measuring the circle radius of $r_e$ on Smith charts similar to that in Fig.~\ref{transmissionspectrum}b for different $\delta\Phi$. Fig.~\ref{efficiency} shows the derived efficiency as a function of generated single photon frequency. We obtained more than 50\% efficiency almost everywhere over the range of frequencies from 6.7 to 9.1~GHz except at one point around $7.25\,$GHz. The efficiency can be affected by non-negligible pure dephasing $\gamma$ and/or non-radiative relaxation $\Gamma_1^{nr}$.
\\

\noindent \textbf{Discussion}\\
In conclusion, we demonstrated an on-chip tuneable on-demand single-microwave-photon source operating with high efficiency over a wide range. The source is expected to be useful for applications including quantum communication, quantum information processing, and sensing.

\noindent \textbf{Acknowledgements}\\
Peng would like to thank Z. R. Lin for help and Y. Kitagawa for preparing Nb wafers. This work was carried out within the project EXL03 MICROPHOTON of the European Metrology Research Programme (EMRP). EMRP is jointly funded by the EMRP participating countries within EURAMET and the European Union. This work was supported by ImPACT Program of Council for Science, Technology and Innovation (Cabinet Office, Government of Japan).

\noindent \textbf{Author contributions}\\
Z.H.P and O.V.A designed, carried out the experiment and analysed the data. Z.H.P and O.V.A wrote the manuscript; and all the authors discussed the data and commented on the manuscript.
 
\noindent \textbf{Additional information}\\
\textbf{Competing financial interests:} The authors declare no competing financial interests. 
\end{document}